\documentclass[aps,pre,twocolumn,amsmath,amsfonts,floatfix,superscriptaddress,longbibliography]{revtex4}
\usepackage{mathrsfs} 
\usepackage{amssymb} 
\usepackage{color} 
\usepackage{bbm} 
\usepackage{multirow}
\usepackage{graphicx}
\usepackage{hyperref}
\usepackage[usenames,dvipsnames]{xcolor}
\definecolor{darkblue}{rgb}{0,0,0.6}

\begin{document}

\title{Brittle yielding of amorphous solids at finite shear rates}

\author{Murari Singh}

\affiliation{Laboratoire Charles Coulomb, 
UMR 5221 CNRS-Universit\'e de Montpellier, Montpellier, France}

\author{Misaki Ozawa}

\affiliation{Laboratoire Charles Coulomb, 
UMR 5221 CNRS-Universit\'e de Montpellier, Montpellier, France}

\affiliation{D\'epartement de Physique, Ecole Normale Sup\'erieure, Paris, France}

\author{Ludovic Berthier}

\affiliation{Laboratoire Charles Coulomb, 
UMR 5221 CNRS-Universit\'e de Montpellier, Montpellier, France}

\affiliation{Department of Chemistry, University of Cambridge, Lensfield Road, Cambridge CB2 1EW, United Kingdom}

\begin{abstract}
Amorphous solids display a ductile to brittle transition as the kinetic stability of the quiescent glass is increased, which leads to a material failure controlled by the sudden emergence of a macroscopic shear band in quasi-static protocols. We numerically study how finite deformation rates influence ductile and brittle yielding behaviors using model glasses in two and three spatial dimensions. We find that a finite shear rate systematically enhances the stress overshoot of poorly-annealed systems, without necessarily producing shear bands. For well-annealed systems, the non-equilibrium discontinuous yielding transition is smeared out by finite shear rates and it is accompanied by the emergence of multiple shear bands that have been also reported in metallic glass experiments. We show that the typical size of the bands and the distance between them increases  algebraically with the inverse shear rate. We provide a dynamic scaling argument for the corresponding lengthscale, based on the competition between the deformation rate and the propagation time of the shear bands.
\end{abstract}

\date{\today}

\maketitle

\section{Introduction}

The mechanical response of amorphous materials such as foams, colloids, and metallic glasses is an active research topic for material science, engineering, and in the context of the physics of disordered systems~\cite{barrat2011heterogeneities,rodney2011modeling,falk2011deformation,bonn2017yield,nicolas2018deformation}. Despite wildly different sizes and interactions of the constituent particles, these diverse materials show surprisingly universal rheological responses under external loadings, such as yielding, plastic rearrangements, avalanches, and shear bands. Concepts and ideas developed in statistical mechanics are thus particularly useful to extract and understand these universal features~\cite{bonn2017yield,nicolas2018deformation}.

Here we focus on the yielding transition of quiescent materials in shear start-up conditions. This problem addresses the basic question of how a given amorphous solid plastically deforms or break when a non-linear mechanical deformation is applied by an external loading. In this setting, two types of yielding transitions can be observed. One type is brittle yielding which is associated with an abrupt failure of the material and corresponds to the apparition of sharp shear bands~\cite{greer2013shear}. The other type is ductile yielding which is accompanied by significant plastic deformations that prevent the emergence of a sharp failure and favor large deformations~\cite{bonn2017yield}. These different yielding behaviors may depend on material properties, preparation protocols, and loading conditions~\cite{schroers2004ductile,lu2003deformation,yang2012size,scholz2019mechanics,popovic2018elastoplastic,shi2005strain,rottler2005unified,kumar2013critical,fan2017effects,moorcroft2011age,vasoya2016notch,jin2017exploring}. In particular, the initial stability of the glass (as controlled by the preparation protocol of the material) directly determines the brittle or ductile nature of yielding. More stable glasses show more brittle yielding, whereas less stable glasses demonstrate ductile behavior~\cite{shi2005strain,kumar2013critical,fan2017effects,vasoya2016notch}.

In the last decade, studies of yielding by the statistical physics community have been largely dedicated to steady state properties after a large accumulated strain. Regarding shear start-up conditions, relatively poorly annealed glasses have been mostly analysed, focusing on plastic rearrangements~\cite{tanguy2006plastic,barrat2011heterogeneities,shrivastav2016yielding}, the formation of shear bands~\cite{shi2005strain,fielding2014shear}, and avalanche statistics~\cite{karmakar2010statistical,salerno2012avalanches,lin2014scaling}. Similar analysis and direct visualisations have also been performed in colloidal glasses~\cite{koumakis2012yielding,ghosh2017direct}.
Many useful concepts have emerged from these intensive investigations, from the definition of soft spots where plastic events successively take place~\cite{manning2011vibrational,patinet2016connecting,cubuk2017structure}, to the localization of plastic events~\cite{dasgupta2013yield} and scaling laws for avalanche statistics~\cite{salerno2012avalanches,lin2014scaling,liu2016driving}. 

By contrast, much less is known about the sharp yielding transition of brittle materials. This problem is however receiving growing attention thanks to the development of novel theoretical approaches~\cite{rainone2015following,wisitsorasak2012strength,urbani2017shear,jaiswal2016mechanical,OBBRT18} and progress in numerical techniques~\cite{ninarello2017models,kapteijns2019fast} that now allow the investigation of brittle yielding in atomistic computer simulations. From a theoretical viewpoint, brittle yielding under quasi-static loading corresponds to a non-equilibrium discontinuous transition. This is described as a spinodal transition in the mean-field limit~\cite{nandi2016spinodals}, potentially avoided in finite dimensions~\cite{OBBRT18}. In addition to these theoretical predictions, molecular simulations in athermal quasi-static shear (AQS) deformation~\cite{maloney2006amorphous} demonstrated that the non-equilibrium discontinuous transition can exist in finite-dimensional models, accompanied by the sudden appearance of a unique system-spanning shear band~\cite{OBBRT18,2dinprepa}.

In the above studies, brittle yielding is described using the language of phase transitions and critical phenomena, but this description applies, strictly speaking, only in the AQS limit. In experiments, several additional factors may play a role and affect yielding, such as thermal fluctuations, spontaneous relaxation, inertia, and a finite deformation rate. In this paper, we deal with the latter and analyse the influence of a finite shear rate, leaving out temperature and inertia in this first effort. The loading rate dependence of yielding and the formation of shear bands is an important topic in material science and engineering~\cite{greer2013shear,antonaglia2014tuned}, as well as soft matter~\cite{amann2013overshoots}. In particular, it has been reported that multiple shear bands appear at a higher strain rate in metallic glass experiments in various rheological conditions, and the density of shear bands increases with increasing $\dot{\gamma}$~\cite{mukai2002effect,schuh2003nanoindentation,sergueeva2004strain,hajlaoui2008strain}. Thus, a computational study about brittle yielding at finite $\dot{\gamma}$ provides useful microscopic insights for both experimental observations at higher strain rates~\cite{wakeda2008multiple}, and for a fundamental understanding of the nature of the yielding transition~\cite{parisi2017shear}.
Physically, we expect that the idealised picture of a single macroscopic shear band being responsible for the failure of the material can not exist at finite shear rate, because it would take an infinite time to create an infinite shear band in an infinite system. The finite timescale introduced by the finite shear rate must compete with the propagation of shear bands. Our main goal is to understand the consequences of this competition and provide a real space picture of yielding. 

In this paper, we perform athermal, overdamped simulations at finite strain rate to shear glasses with a broad range of initial stabilities, in order to characterise the relevant timescales and lengthscales associated with brittle yielding at finite strain rate. By measuring the stress-strain curve and associated susceptibilities, we find that the discontinuous yielding transition observed in the AQS simulations is smeared out in finite strain rate simulation, when $\dot{\gamma}$ is high and the system size $N$ is large. Larger samples require slower $\dot{\gamma}$ to display brittle yielding with a single system spanning shear band. If $\dot{\gamma}$ is large for a given system size, we instead observe that multiple shear bands emerge, as reported in metallic glass experiments. We then extract a typical lengthscale $\xi$ characterizing the spatial pattern of shear bands for a given $\dot{\gamma}$. We find that $\xi$ scales as $\xi  \propto \dot{\gamma}^{-\alpha}$, where $\alpha \approx 0.4$ for two-dimensional stable glasses. Thus, the lengthscale $\xi$ diverges in the AQS limit, for sufficiently stable glasses. We argue that the observed scaling behavior can be understood as the competition between the deformation rate and the timescale associated with shear band formation.

This manuscript is organized as follows. In Sec.~\ref{sec:methods}, we describe the numerical methods. In Sec.~\ref{sec:macro}, we present the macroscopic rheological properties of glasses prepared with different initial stabilities to expose the basic differences between ductile and brittle yielding at finite strain rates. Section~\ref{sec:transition} describes the effect of the finite strain rate on the nature of the yielding transition. The relevant lengthscale for the yielding transition is visualized and quantified in Sec.~\ref{sec:real}. Finally we discuss and conclude our results in Sec.~\ref{sec:discussions}.

\section{Numerical Methods}
\label{sec:methods}

\subsection{Simulation models}

We simulate size polydisperse systems of $N$ particles 
in cubic and square box of length $L$ in three (3D) and two (2D) dimensions using periodic boundary conditions. The pair interaction between particles $i$ and $j$ separated by a distance $r_{ij}$ is a soft repulsive potential,
\begin{eqnarray}
\label{eq:pot}
u(r_{ij},d_{ij}) / \epsilon &=& \left(\frac{d_{ij}}{r_{ij}}\right)^{12} + c_0  + c_2 \left(\frac{r_{ij}}{d_{ij}}\right)^{2} + c_4 \left(\frac{r_{ij}}{d_{ij}}\right)^{4},  \nonumber \\ 
d_{ij} &=& \frac{d_i+d_j}{2}\left(1-0.2|d_i-d_j|\right), \nonumber
\end{eqnarray}
where $r_{ij}$ is the distance between particles $i$ and $j$, $d_i$ is the diameter of the particle $i$, and $\epsilon$ is the energy scale. The set of parameters, $c_0$, $c_2$, and $c_4$, are adjusted so that the potential and its first and second derivatives vanish at the cutoff distance $r_{{\rm cut}, ij}=1.25d_{ij}$. The particle diameters are drawn randomly from a continuous size distribution $P(d)=A/d^3$ in the range $[d_{\rm min},d_{\rm max}]$, where $A$ is normalizing constant. We use parameters such that  $d_{\rm min}/d_{\rm max}=0.45$ and the average size diameter is $\overline{d}=1.0$. We  perform  simulations at constant number density $\rho=1.02$ for 3D, and $\rho=1$ for 2D, using different system sizes $N \in [1500,96000]$ in 3D and $N=64000$ in 2D.

To prepare the glassy samples to be sheared at temperature $T=0$, we first equilibrate the system at some finite temperature, $T_{\rm ini}$, with the help of an efficient swap Monte Carlo method~\cite{ninarello2017models}. The equilibrium configurations are then instantaneously quenched at $T=0$ using the conjugate gradient method~\cite{nocedal2006numerical}. We produce glassy samples using initial temperatures $T_{\rm ini}\in[0.062,0.200]$, which offers a broad range of kinetica stability. In 2D, the initial preparation temperatures are $T_{\rm ini}\in[0.035,0.200]$~\cite{berthier2019zero}.
For these temperature ranges, we can cover in both 3D and 2D the range of behaviour between brittle and ductile when AQS simulations are used~\cite{OBBRT18,2dinprepa}.

\subsection{Equations of motion}

Our goal is to analyse the effect of a finite shear rate on the brittle yielding transition observed in AQS conditions reported in Ref.~\cite{OBBRT18}. To avoid adding too many ingredients at once, we study the dynamics at zero temperature in the absence of inertia. To this end, we perform molecular dynamics simulations using overdamped Langevin equations of motion at $T=0$~\cite{durian1997bubble}. We impose a simple shear flow in the $\hat{x}$ direction, where $\hat{x}$ is the unit vector along the $x$ axis, and solve the following equations of motion,
\begin{equation}
\zeta \left(  \frac{\mathrm{d} {\bf r}_i}{\mathrm{d} t} -\dot{\gamma}y_i \hat{x}   \right)= - \sum_{j < k} \frac{\partial u(r_{jk}, d_{jk})}{\partial {\bf r}_i},
\label{eq:shearrate}
\end{equation}
where $\zeta$ is the viscous damping coefficient, ${\bf r}_i$ and $y_i$ represent the position and its $y$ component of a particle. We use Lees-Edwards boundary conditions to perform simulations at a finite shear strain rate $\dot{\gamma}$~\cite{at89}.

In the absence of thermal fluctuations, the natural microscopic timescale is given by $\tau_0 = \zeta \overline{d}^2/\epsilon$ which controls the viscous dissipation of the system. Length, time, and energy are expressed in units of $\overline{d}$, $\tau_0$, and $\epsilon$, respectively. To integrate the equations of motion in Eq.~(\ref{eq:shearrate}), we employ the Runge-Kutta method of order 4 with a time step $\Delta t = 0.005$ and the Euler method with a time step $\Delta t = 0.001$~\cite{fs02}. We confirmed that these two methods produce identical results. We compute the $xy$ component of the stress, $\sigma$, using the Irving-Kirkwood formula.

Additionnally, we perform strain-controlled athermal quasi-static
shear (AQS) deformation using Lees-Edwards boundary conditions~\cite{ML06}, to complement data obtained for the same model in Ref.~\cite{OBBRT18}. The AQS shear method consists of a succession of tiny uniform shear deformation with a step size of $\Delta \gamma=10^{-4}$ followed by energy minimization via the conjugate gradient method.

By definition, the finite strain rate simulations described by Eq.~(\ref{eq:shearrate}) should produce results identical to the AQS simulations in the limit $\dot{\gamma} \to 0$. Therefore, Eq.~(\ref{eq:shearrate}) is the simplest and most natural extension of the AQS study of brittle yielding to finite strain rates, which introduces only one additional control parameter, $\dot{\gamma}$. In future work, it would be interesting to study the effect of temperature and of inertia on this phenomenon, which also introduce additional timescales in the problem.

\section{Macroscopic Rheology}

\label{sec:macro}

\subsection{Steady State flow curve}

\begin{figure}
\centering
\includegraphics[width=8cm]{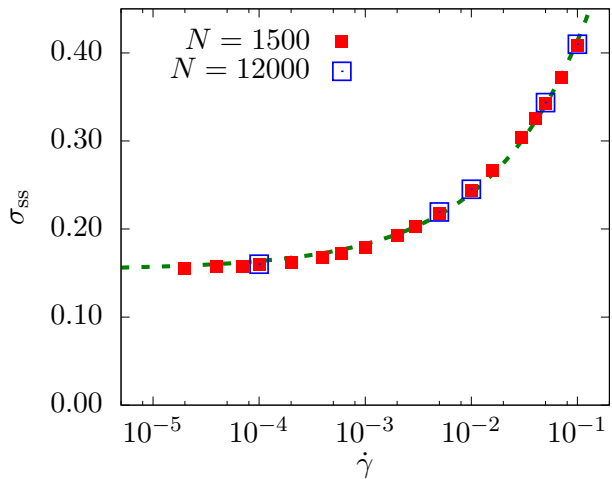}
\caption{Average shear stress in the steady state regime as a function of applied shear strain rate for the 3D with $N=1500$ and $12000$. The green dashed curve represents the Herschel-Bulkley law with a yield stress independently measured in the AQS limit at $\dot{\gamma}=0$.}
\label{fig:sigmaflow}
\end{figure}

Before showing results for the shear start-up setting, we present the steady state flow curve to illustrate the range of $\dot{\gamma}$ that we impose, and the basic rheological properties of our numerical models at finite $\dot{\gamma}$ in the steady state, where no shear band is present.

In Fig.~\ref{fig:sigmaflow}, we present the steady state 
flow curve for the 3D system and two system sizes, $N=1500$ and $N=12000$. The average of the shear stress in the steady-state,  $\sigma_{\rm ss}$, is obtained by averaging over many configurations for strain larger than 1000$\%$ and over different samples.  We do not observe finite size effects, at least down to $\dot{\gamma}=10^{-4}$. 

We independently measure the shear stress $\sigma_{\rm ss}^{\rm AQS}=0.154$ in the steady state for the AQS condition. We substitute this value in the Herschel Bulkley equation~\cite{HB1926}, $\sigma_{\rm ss}=\sigma_{\rm ss}^{\rm AQS} + B \dot{\gamma}^n$, where $B$ is a prefactor, and find that this phenomenological equation describes our data very well with the exponent $n=0.515$. This value seems consistent with several earlier studies~\cite{ikeda2012unified,vasisht2018rate,pinaki2018arXiv}. 

In the steady state, we do not observe any instability, such as ordering along the shear direction, or shear localization. Besides, the obtained flow curve in Fig.~\ref{fig:sigmaflow} does not present a non-monotonic behaviour. Thus, the system under study does not satisfy any known condition to produce permanent shear bands in the steady state~\cite{bonn2017yield,martens2012spontaneous,fielding2014shear}.
In other words, the shear bands observed in our study in the shear start-up setting are inherently a transient phenomenon whose origin can directly be related to the nature of the initial configurations~\cite{moorcroft2011age}. 

\subsection{Shear start-up}

We now focus on the macroscopic stress-strain curves obtained in the shear start-up setting. We prepare zero-temperature glasses at various depth in their energy landscape quantified by the preparation temperature $T_{\rm ini}$ and apply a finite shear rate at time $t=0$. For each $T_{\rm ini}$ we average the results over independent glass configurations to increase the statistics of the data to obtain the evolution of the average shear stress, denoted by $\langle \sigma \rangle$, as a function of the deformation $\gamma = \dot{\gamma} t$ since time $t=0$. We present the results for poorly annealed glasses prepared at high temperature, $T_{\rm ini}=0.200$, and for very stable glasses prepared at low temperature, $T_{\rm ini}=0.062$. 

\begin{figure}
\centering
\includegraphics[scale=1]{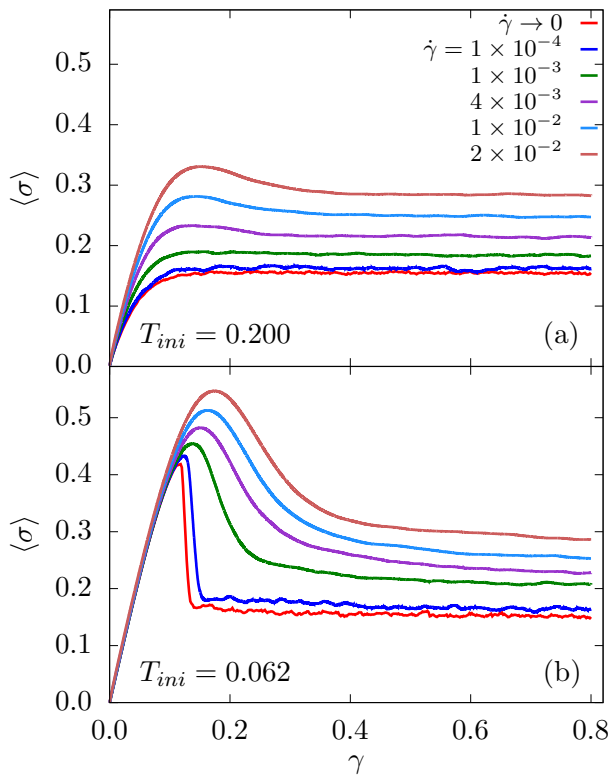}
\caption{Average stress strain curves for 3D glasses for $N=12000$ (a) at high initial preparation temperature $T_{\rm ini}=0.200$, and (b) at low initial preparation temperature $T_{\rm ini}=0.062$. The data denoted $\dot{\gamma} \to 0$ are measured using AQS simulations.}
\label{fig:stressstrainRate}
\end{figure}

In Fig.~\ref{fig:stressstrainRate}(a), we report the results for poorly annealed glasses. First, we show that in the AQS simulation, $\dot{\gamma} \to 0$, the system shows a completely monotonic crossover across yielding and reaches steady state without any stress overshoot, consistent with a very ductile behaviour. When a finite strain rate is applied, deviations from the AQS results are clearly observed. As $\dot{\gamma}$ is increased, we observe that up to a strain rate $\dot{\gamma} \approx 10^{-3}$, the system shows a qualitatively similar monotonic crossover to yielding, akin to the AQS conditions. Incrementing the strain rate further, $\dot{\gamma}\geq 4 \times 10^{-3}$, the system starts to present a stress overshoot during yielding~\cite{varnik2004study,tsamados2010plasticity}. 
The same trend can be observed for different system sizes from $N=1500$ to $96000$. 

It has been theoretically argued that the presence of the stress overshoot before reaching the steady state is associated with the emergence of shear banding~\cite{MCF11}. However, a careful analysis of the non-affine displacement field for these systems does not reveal any sign of a transient shear band in our simulations. Whereas shear bands may well appear at even larger system sizes~\cite{vasisht2018permanent}, we feel that these systems are too ductile to show any interesting shear localisation at such large shear rates. 

In Fig.~\ref{fig:stressstrainRate}(b), we show the results obtained for very stable glasses. We find that all samples display a sharp shear band at low shear rates, which can be either horizontal or vertical. We average the stress strain curves over the samples showing horizontal shear bands, in order to remove the small stress growth observed after yielding when a vertical shear band is present~\cite{kapteijns2019fast}. In this case, we observe that in the AQS limit the system shows a discontinuous stress drop after stress overshoot, as reported previously~\cite{OBBRT18}. At finite but low strain rate, $\dot{\gamma}=10^{-4}$, we observe that the average stress strain curve shows a trend very similar to the AQS results, with a very slight change of the slope $d \langle \sigma \rangle/ d \gamma$ precisely at yielding as compared to the AQS limit. In that case, we also observe a system spanning shear band, as discussed in more details below in Sec.\ref{sec:real}.
When the strain rate is increased further, $\dot{\gamma} \geq 10^{-3}$, the sharp stress drop at yielding is smeared out, and the corresponding slope is also systematically decreased. Concomitantly, the formation of a shear band is also altered, as described below in Sec.\ref{sec:real}.

For stable glasses, reaching the steady-state requires straining the sample for extremely large values of $\gamma$. For example, $\gamma > 3.0$ is needed to reach the steady state at $\dot{\gamma}=10^{-1}$ for $N=12000$. Besides, the amount of strain to reach the steady state increases with increasing the system size or decreasing $\dot{\gamma}$. At the slowest shear rate limit, $\dot{\gamma} \to 0$, $\gamma=10.0$ is not enough at all to reach the steady state for $N=12000$ systems. Thus, the sheared glassy states obtained immediately after the stress overshoot in Fig.~\ref{fig:stressstrainRate}(b), say $0.6 \leq \gamma \leq 0.8$, are not yet typical of the steady state, even though the stress seems to have reached a stationary strain dependence. The energy is a more suitable observable to detect whether a steady state has been reached.

\section{Yielding Transition}

\label{sec:transition}

Having exposed the basic phenomenology at finite strain rates in the previous section, we now focus on the effect of a finite $\dot{\gamma}$ on the non-equilibrium first-order transition observed in stable glasses in AQS simulations~\cite{OBBRT18}. 

\subsection{Finite size scaling analysis}

\begin{figure}
\includegraphics[scale=1]{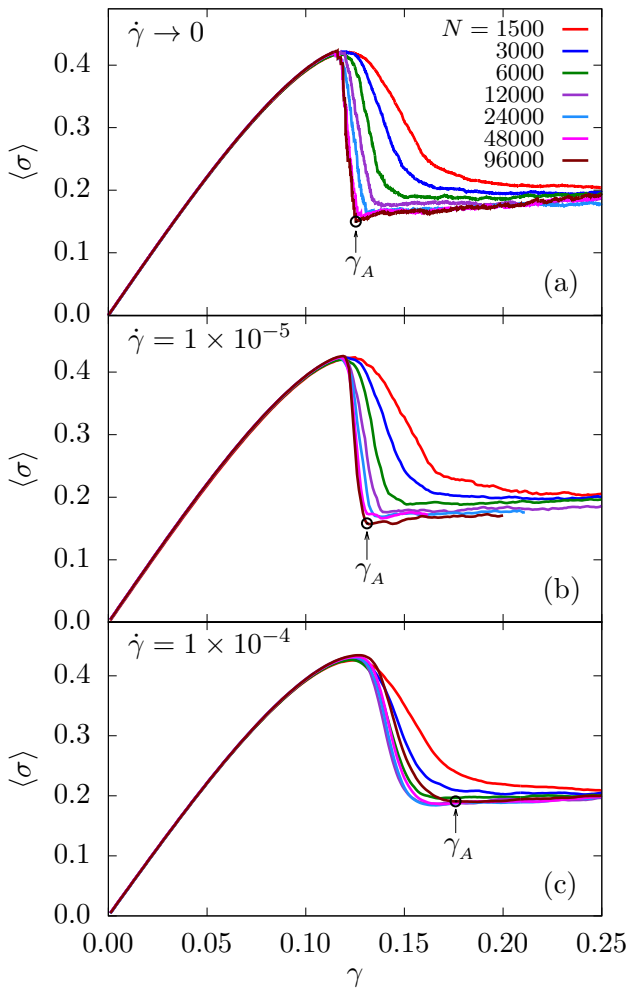}
\caption{Average stress strain curves for 3D AQS simulations (a), $\dot{\gamma}=10^{-5}$ (b), and $\dot{\gamma}=10^{-4}$ (c) and different system sizes. The preparation temperature is $T_{\rm ini}=0.062$. The vertical arrow indicates the location of $\gamma_{\rm A}$ used in Sec.~\ref{sec:real} to study shear band formation.}
\label{fig:avgStressStrain}
\end{figure}

In Fig.~\ref{fig:avgStressStrain}(a), we display stress strain curves for a stable glass in AQS simulations, varying the system size $N$. These data were first shown in Ref.~\cite{OBBRT18}. The slope of these curves just after the stress overshoot becomes sharper with increasing system size, and the stress drop becomes a genuine discontinuity in the thermodynamic limit.

When a finite strain rate $\dot{\gamma} = 10^{-5}$ is applied for the same system sizes, see Fig.~\ref{fig:avgStressStrain}(b), the stress drop at yielding still gets increasingly sharper with system size, at least up to $N=96000$. For these system sizes, then, we observe only little difference between this small shear rate and the AQS limit. 

However, when the strain rate is increased further, $\dot{\gamma} = 10^{-4}$, we observe that for $N \geq 12000$, the stress strain curves no longer evolve and the stress drop does not become sharper at larger $N$, see Fig.~\ref{fig:avgStressStrain}(c). 

In summary, we find that for a finite strain rate, the sharp stress drop seen in the AQS limit initially gets sharper with increasing the system size, but there seems to exist a finite $N$ above which it saturates. This crossover system size becomes larger for smaller shear rate, and presumably it diverges as $\dot{\gamma} \to 0$, so that the AQS discontinuous limit is recovered. These observations suggest that non-equilibrium first-order transition seen in the AQS simulation is now smeared out by the finite timescale introduced by the shear rate. 

\subsection{Stress fluctuations and susceptibilities}  

\begin{figure}
\includegraphics[scale = 1]{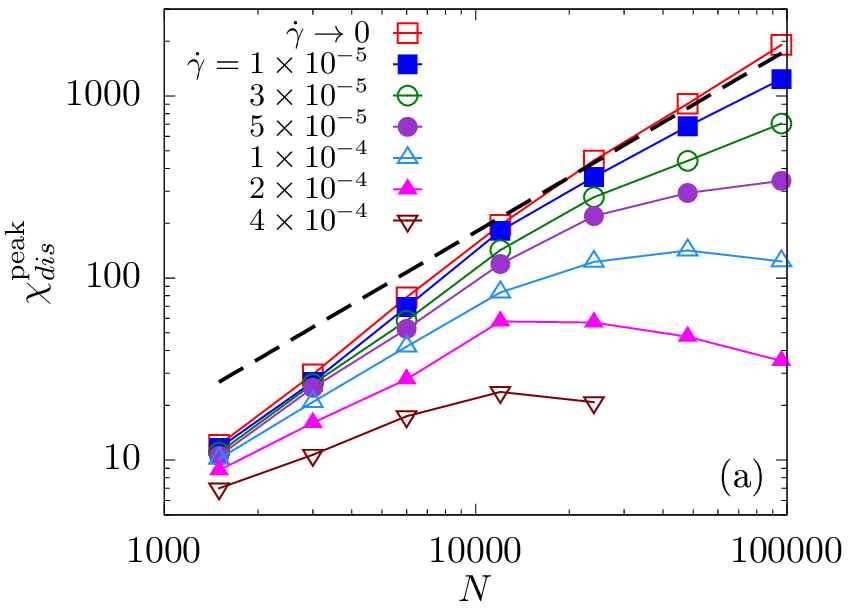}\\
\includegraphics[scale = 0.98]{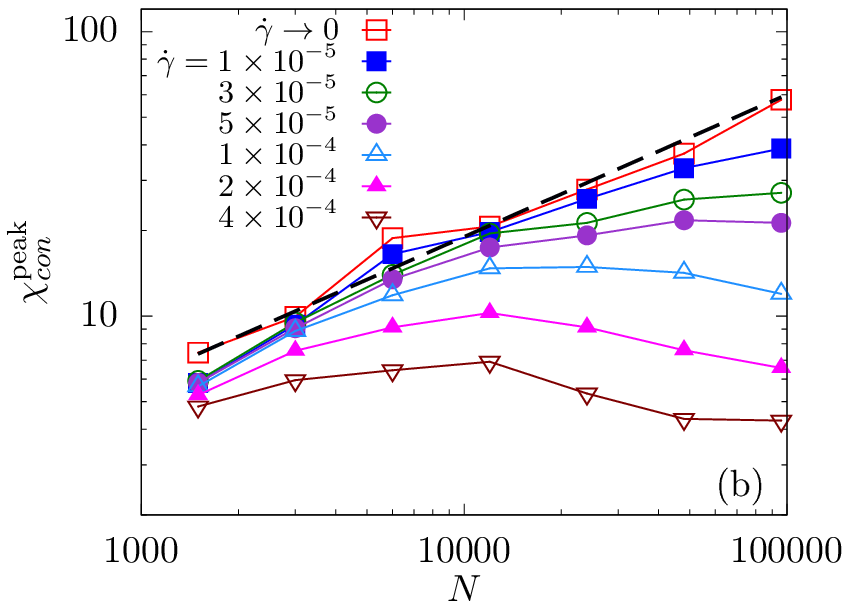}\\ 
\caption{The peak values of the disconnected (a) and connected (b) susceptibilities, $\chi_{\rm dis}^{\rm peak}$ and $\chi_{\rm con}^{\rm peak}$, as a function of the system size $N$ for 3D stable glasses with $T_{\rm ini}=0.062$. The dashed lines correspond to the scaling behaviors observed in the AQS limit, $\chi_{\rm dis}^{\rm peak} \propto N$ and $\chi_{\rm con}^{\rm peak} \propto N^{1/2}$. Departure from the AQS limit arises at smaller $N$ when $\dot{\gamma}$ increases, indirectly revealing a crossover lengthscale $\xi$ controlled by the shear rate $\dot{\gamma}$.}
\label{fig:susc}
\end{figure}

In the AQS limit, the brittle yielding transition is most transparently revealed via the analysis of the stress fluctuations, which are efficiently quantified by two susceptibilities that we now discuss. 

We first define the connected susceptibility, 
\begin{equation}
\chi_{\rm con}= -d \langle \sigma \rangle /d \gamma,
\end{equation} 
which can be directly measured from the dervative of the average stress strain curves shown in Fig.~\ref{fig:avgStressStrain}. The second quantity is the disconnected susceptibility, 
\begin{equation}
\chi_{\rm dis}= N(\langle \sigma ^2 \rangle - \langle \sigma \rangle ^2),
\end{equation} 
which quantifies the sample to sample fluctuations of the shear stress at a given strain $\gamma$. Both these susceptibilities exhibit a pronounced peak near yielding, and we define $\chi^{\mathrm{peak}}_{\rm con}$ and $\chi^{\mathrm{peak}}_{\rm dis}$ as the amplitude of these peaks. These amplitudes thus depend on the preparation temperature $T_{\rm ini}$, on the system size $N$, and on the applied shear rate $\dot{\gamma}$. 

In Fig.~\ref{fig:susc}, we show the evolution of these peak values for 3D stable glasses with $T_{\rm ini}=0.062$. In the AQS limit, the system size dependence of the susceptibilities is well understood~\cite{OBBRT18}. They both diverge as a power law of the system size, $\propto N^\delta$ with 
$\delta=1$ and $\delta = 1/2$ for the connected and disconnected susceptibilities, respectively. These divergences at $\dot{\gamma} \to 0$ reflect the existence of sharp non-equilibrium first-order transition in the thermodynamic limit~\cite{OBBRT18}. 

When we apply a low finite $\dot{\gamma}$, at smaller $N$, $\chi^{\mathrm{peak}}_{\rm dis}$ and  $\chi^{\mathrm{peak}}_{\rm con}$ still follow the AQS behaviour for small enough $N$. However, they depart from the AQS behaviour for larger $N$ and the divergence with $N$ is eventually avoided and replaced by a saturation of the fluctuations to a finite value. The deviations from the AQS limit become stronger with increasing the shear rate. These results confirm the existence of a crossover system size, $N^*(\dot{\gamma})$, below which the AQS behaviour is observed, but above which the divergence of the stress fluctuations is cutoff. This crossover system size $N^*(\dot{\gamma})$ becomes larger for smaller shear rate, and it diverges in the AQS limit $\dot{\gamma} \to 0$. 

These observations imply that there is a direct connection between the timescale imposed by the shear rate and a typical lengthscale $\xi=\xi(\dot{\gamma})$ characterizing the yielding transition. Since brittle yielding is associated with a single system spanning shear band, the crossover lengthscale revealed by the above analysis suggests the emergence of a characteristic lengthscale associated with shear band formation at finite shear rates, so that a sharp behaviour similar to AQS physics is observed when $L \ll \xi$, whereas a new physical regime is entered when $L \gg \xi$. This conclusion suggests that a direct visualisation of the real space deformations of yielding is needed, which is the topic of the next section. 
 
\section{A lengthscale associated with shear band formation}

\label{sec:real}

The purpose of this section is to analyse in real space how the sharp yielding transition observed in AQS conditions for stable glasses is modified when using a finite shear rate.
 
To this end, we spatially resolve the plastic activity by measuring the   accumulated non-affine displacement for each particle. We follow the standard method introduced by Falk and Langer to compute the quantity $D^2_{\rm min}$, which provides the local deviation of the particle displacement from an affine deformation~\cite{FL98}. At a given strain $\gamma$, we compute the deformation with respect to the unstrained sample at $\gamma=0$. We also follow them~\cite{FL98} regarding the definition of neighboring particles and use a cut-off radius $r_{\rm cut}=2.5$.
 
\subsection{Visualisation in 3D}

To best visualize the plastic activity in the strained samples, we need to choose a strain value very close to yielding, right after the stress drop that is observed in the macroscopic stress-strain curves. While this choice is easily made in the AQS limit where the stress drop in stable glasses is essentially instantaneous, this is less obvious at finite $\dot{\gamma}$. Typical choices are shown in Fig.~\ref{fig:avgStressStrain}.
As seen in Fig.~\ref{fig:avgStressStrain}, for a given system size, $\gamma_{\rm A}$ should increase as $\dot{\gamma}$ increases. For example for $N=96000$, in the AQS limit this value is $\gamma_{\rm A} \approx 0.125$, whereas at the strain rate $\dot{\gamma} =10^{-5}$ and $\dot{\gamma}=10^{-4}$ take $\gamma_{\rm A} \approx 0.130$ and $\gamma_{\rm A} \approx 0.176$, respectively. All the images shown below are for the largest system simulated in 3D, $N=96000$.

\begin{figure}
\includegraphics[width=8cm]{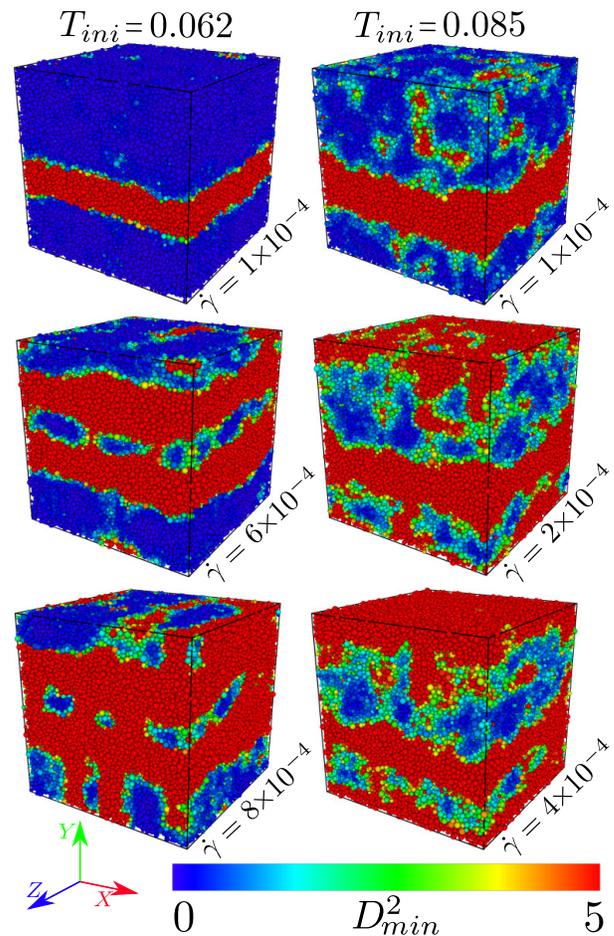}
\caption{Snapshots of the non-affine $D^2_{\rm min}$ field immediately after yielding (at the strain $\gamma_{\rm A}$ shown in Fig.~\ref{fig:avgStressStrain}) for two  different $T_{\rm ini}$ and various shear rates $\dot{\gamma}$ for the 3D system with $N=96000$. An increasing number of shear bands is observed as the strain rate increases.}
\label{fig:3dsnaps}
\end{figure}

We summarize our observations in Fig.~\ref{fig:3dsnaps}, which shows snapshots for different $\dot{\gamma}$ at the corresponding $\gamma_{\rm A}$ for glasses prepared initially at $T_{\rm ini}=0.062$ and $T_{\rm ini}=0.085$. 
Both preparation temperatures are below the brittle-to-ductile critical temperature of the AQS condition, $T_{\rm ini, c} \approx 0.095$, and thus show sharp discontinuous stress drops in AQS simulations~\cite{OBBRT18}.
  
First, we discuss the results for the best annealed sample ($T_{\rm ini}=0.062$). 
Whereas we have observed smearing out of the sharp stress drop at a rate of $\dot{\gamma}=1\times10^{-4}$ (see Fig.~\ref{fig:avgStressStrain}), the system still forms a well-defined single shear band right after yielding. As we increase the strain rate to $\dot{\gamma}=6\times 10^{-4}$, two shear bands are typically observed, with additional smaller plastic events seen elsewhere in the system. Increasing even further the strain rate to $\dot{\gamma}=8\times 10^{-4}$, we now observe multiple shear bands in both horizontal and vertical directions. A qualitatively similar behaviour is observed for $T_{\rm ini}=0.085$, but many more plastic events are already present at the smallest shear rate $\dot{\gamma}=1\times10^{-4}$, which coexist with a macroscopic shear band. Again, increasing the shear rate results in multiple shear bands that are less and less well-resolved. 

The emergence of multiple shear bands at high loading rates has also been reported in metallic glass experiments in various loading conditions, compressive~\cite{mukai2002effect,antonaglia2014tuned}, tensile~\cite{sergueeva2004strain,hajlaoui2008strain}, and nanoindentation~\cite{schuh2003nanoindentation} deformation tests, as well as molecular dynamics simulations~\cite{wakeda2008multiple}. These observations, together with our numerical results, could be interpreted as follows. When $\dot{\gamma}$ is finite, the system does not have enough time to develop a single system spanning shear band. Instead, the system responds to the applied strain by independently forming several shear bands at various locations in the material~\cite{hajlaoui2008strain}. 

\subsection{Visualisation in 2D}

\begin{figure*}
\centering
\includegraphics[width=18cm]{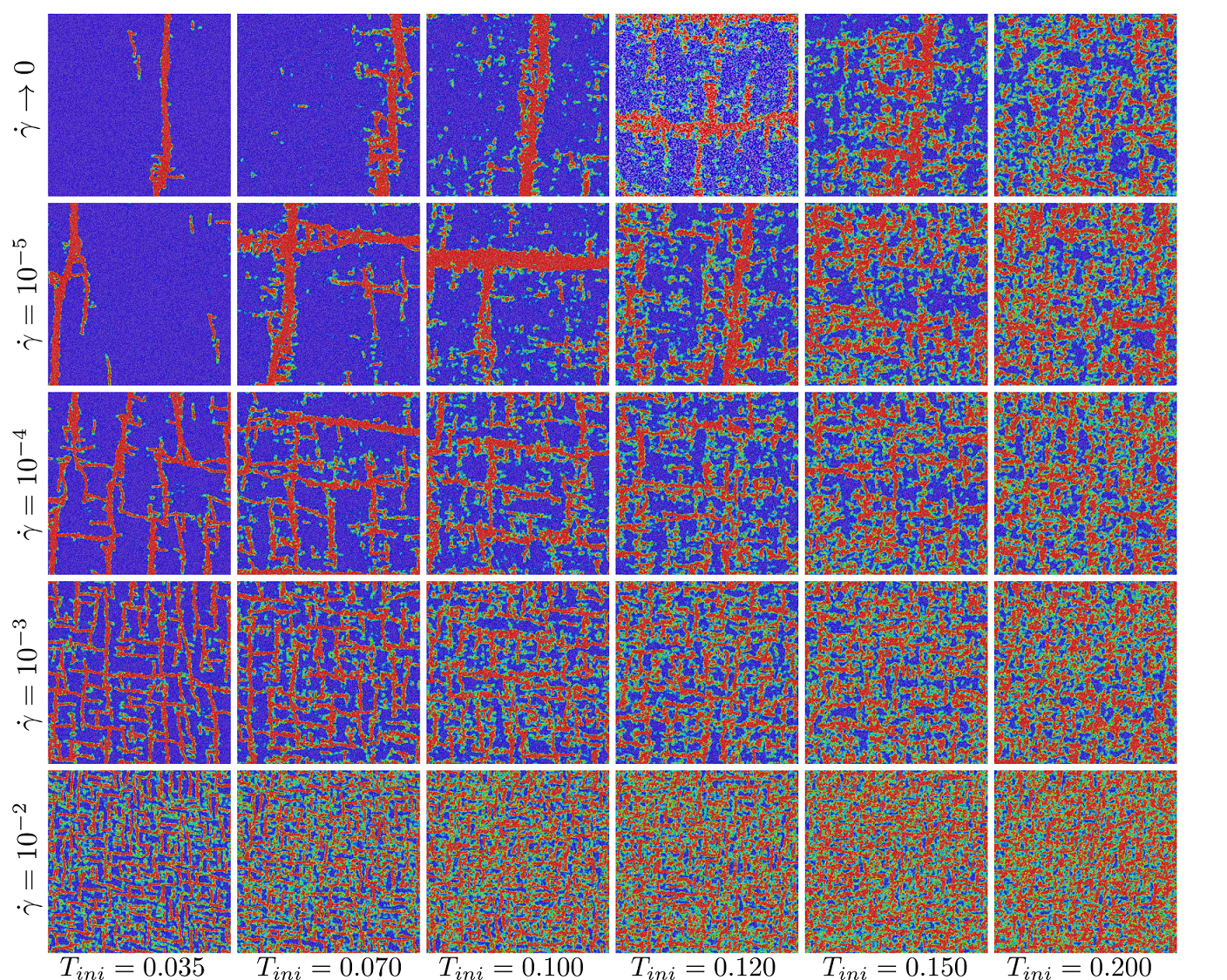}
\caption{
Snapshots of the non-affine $D^2_{\rm min}$ field immediately after yielding (at the strain $\gamma_{\rm A}$) for several $T_{\rm ini}$ and various shear rates $\dot{\gamma}$ for the 2D system with $N=64000$ ($L=253$). An increasing number of shear bands is formed as the strain rate increases, which form a checkerboard pattern for low $T_{\rm ini}$ with characteristic lengthscale $\langle \xi \rangle$.}
\label{fig:2dsnaps}
\end{figure*}

From the above real space observations in 3D, we concluded that an increasing strain rate produces multiple shear bands, suggesting that a typical finite distance, $\xi$, between shear bands emerges at finite $\dot{\gamma}$, and decreases at larger $\dot{\gamma}$. We postulate that this is a relevant lengthscale for the yielding transition, that we wish to characterize further. 

It is however difficult to quantify this lenth scale from the simulations shown in Fig.~\ref{fig:3dsnaps} because the system size remains too small, despite the fact that we use $N=96000$ particles, which corresponds to a linear system size $L=45.5$. To overcome this difficulty, we perform a similar analysis in 2D systems. This allows us to access larger linear sizes ($L=253$ for $N=64000$), and thus to quantitatively determine how the lengthscale $\xi$ varies with $\dot{\gamma}$. The AQS limit for the 2D model is studied more carefully in Ref.~\cite{2dinprepa}. 

As in 3D, we vary the initial stability of the glass samples by changing  $T_{\rm ini}$ over a considerable range, from very stable glasses to poorly annealed materials in order to highlight the nature of the measured lengthscale $\xi$. In particular, we wish to discriminate $\xi$ from previously reported lengthscales in the literature, discussed in the steady state or poorly-annealed materials~\cite{lemaitre2009rate,KLPZ10}. 

We again analyze the configurations of the 2D systems at the strain value immediately after the stress overshoot $\gamma_{\rm A}$. For a given $\dot{\gamma}$, $\gamma_{\rm A}$ depends very little on $T_{\rm ini}$ (within 5\%) 
and therefore we choose the same $\gamma_{\rm A}$ irrespective of $T_{\rm ini}$. 
We use $\gamma_{\rm A}=0.070$, $0.107$, $0.125$, $0.150$, and $0.190$, for $\dot{\gamma} \to 0$ (AQS), $\dot{\gamma}=10^{-5}$, $10^{-4}$, $10^{-3}$, and $10^{-2}$, respectively.

We summarize our results in the snapshots shown in Fig.~\ref{fig:2dsnaps} where several strain rates and preparation temperatures are shown. 
Starting with the AQS limit (top row), we observe homogeneous plastic activity for poorly annealed systems, and a gradual emergence of shear localisation 
as $T_{\rm ini}$ is decreased, followed by a single sharp shear band at low $T_{\rm ini}$. This evolution mirrors the physics observed in 3D AQS simulations~\cite{OBBRT18}, and its nature is discussed further in Ref.~\cite{2dinprepa}.
We find that there is a critical preparation temperature $T_{\rm ini,c} \approx 0.1$ separating brittle and ductile yielding behaviors also in two dimensions~\cite{2dinprepa}. For $T_{\rm ini}$ close to $T_{\rm ini,c}$, the snapshots reveal a combination of randomly distributed plastic events and a system-spanning shear band. 

When $\dot{\gamma}$ is increased from $\dot{\gamma}=10^{-5}$ to $\dot{\gamma}=10^{-2}$, for poorly annealed samples at $T_{\rm ini}=0.200$, the plastic rearrangements continue to fill the entire sample and remain  homogeneously distributed in space. The shear rate plays only a minor role in these snapshots, the contrast between regions with large and small non-affine displacements become less pronounced as $\dot{\gamma}$ increases. Despite the existence of a stress drop in the average stress-strain curves, these samples do not display shear localisation. 
 
On the other hand, for stable glass samples at $T_{\rm ini}=0.035$, multiple shear bands appear in both horizontal and vertical directions, as seen for instance for $\dot{\gamma}=10^{-4}$. This result is similar to the above observations in 3D. Although less striking, it also appears that the width of each shear band becomes thinner at larger shear rate~\cite{manning2009rate}.
As $\dot{\gamma}$ is increased further for the stable glasses, the density of shear bands increases, or, equivalently, the typical distance between two shear bands decreases. When several shear bands appear inside the system they form a sort of `checkerboard' structure (see for instance $\dot{\gamma}=10^{-3}$ and $\dot{\gamma}=10^{-2}$). Finally, the 
samples at intermediate temperatures, $T_{\rm ini}=0.070-0.150$, appear as a superposition of the two extreme cases ($T_{\rm ini}=0.035$ and $0.200$), with homogeneously spread plastic activity superposed to a checkerboard pattern.
 
\subsection{A shear rate dependent lengthscale for shear banding}

To quantify the typical distance between two shear bands which would be the relevant lengthscale associated with shear band formation in the system, we first compute $D^2_{\rm min}$ for each particle. Particles with large  $D^2_{\rm min}$ typically belong to the shear band, whereas low plastic activity is revealed by a small $D^2_{\rm min}$. This quantity takes however continuous values. We first transform it into a binary variable to more clearly distinghuish the shear bands from the rest of the system. We use a threshold value $D^2_{\rm min}=2.0$, below which we consider the region as being outside the shear band. This binary field now clearly specifies the interface separating the two regions. We checked that transforming the snapshots in Fig.~\ref{fig:2dsnaps} using the binary field leaves the images essentially unaffected. 

This binary information can then be used to compute the typical lengthscale between the shear bands, by measuring the chord length distribution, recording distances between the shear bands. We define a chord by two consecutive intersections of a straight line with the non-shear band region present in the system~\cite{testard2011influence}. To gather statistics we draw many straight lines in both $x$ and $y$ directions, taking care of the Lees-Edwards boundary conditions, and measure the length $\xi$ of each chord along each straight line. We therefore measure for each configuration a distribution of chord lengths, $P(\xi)$. We can then define an average distance between shear bands as the first moment of this distribution, 
\begin{equation}
\langle \xi \rangle = \int_0^\infty \mathrm{d}\xi^\prime P(\xi^\prime) \xi^\prime.
\end{equation}
By construction, $\langle \xi \rangle$ should thus quantitatively represent the typical size of the checkerboard patterns shown in Fig.~\ref{fig:2dsnaps}.
This lenth scale characterizing yielding has a very different nature from lengthscales reported previously, such as the typical distance between plastic events measured in the steady state~\cite{lemaitre2009rate,KLPZ10}. 

\begin{figure}
\includegraphics[scale = 1]{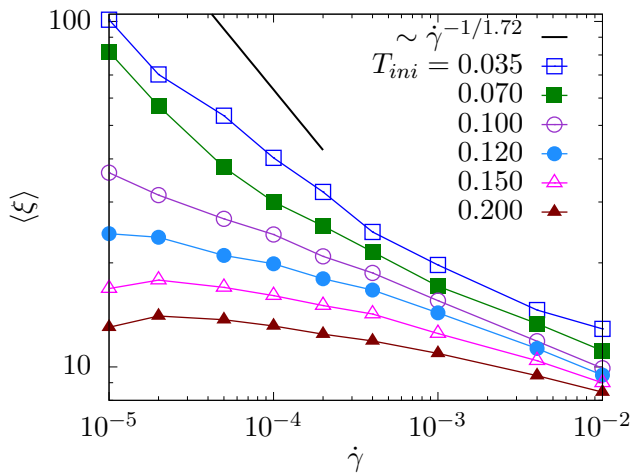}
\caption{Evolution of the typical distance between shear bands with the shear rate for the 2D model at various value of $T_{\rm ini}$. The solid straight line corresponds to the scaling predicted by Eq.~(\ref{eq:scaling}) using the exponent $\beta = 1.72$ measured in Fig.~\ref{fig:relaxation_time},
which should hold for low $T_{\rm ini}$ and small $\dot{\gamma}$.}
\label{fig:lengthXi}
\end{figure}

We show in Fig.~\ref{fig:lengthXi} the evolution of the measured $\langle \xi \rangle$ as a function of $\dot{\gamma}$ for the entire range of $T_{\rm ini}$ analysed in 2D. Stable glasses at $T_{\rm ini}=0.035$ and $0.070$ show monotonic growth of $\langle \xi \rangle$ with decreasing $\dot{\gamma}$, as expected from the direct visualization in Fig.~\ref{fig:2dsnaps}.
For these stable glasses, $\langle \xi \rangle$ seems to grow algebraically at low shear rate, 
\begin{equation}
\langle \xi \rangle \propto \dot{\gamma}^{-\alpha},
\label{eq:alpha}
\end{equation}
with $\alpha \approx 0.4$, suggesting that the distance between shear bands indeed diverges in the AQS limit $\dot{\gamma} \to 0$. This divergence implies that a single shear band exists in fully AQS simulations in the thermodynamic limit.

On the other hand, for less well-annealed glasses, especially $T_{\rm ini}=0.150$ and $0.200$, the lengthscale $\langle \xi \rangle$ saturates  at small $\dot{\gamma}$ to a finite value, suggesting that these systems exhibit a homogeneous spatial distribution of plastic events at large scale, $L \gg \langle \xi \rangle$. Interestingly, glass samples prepared near the critical point, $T_{\rm ini,c} \approx 0.1$ seem to also exhibit a power law divergence with $\dot{\gamma}$, albeit with a different exponent $\alpha$. It would be interesting to relate this exponent to the criticality discussed in Ref.~\cite{OBBRT18}. We also anticipate that $\langle \xi \rangle (\dot{\gamma}, T_{\rm ini})$ may obey a form of critical scaling as a function of the distance $\Delta T  = |T_{\rm ini} - T_{{\rm ini},c}|$. Our data remain insufficient to study these critical behaviours.  

More broadly, these results suggest that the lengthscale $\langle \xi \rangle$ introduced above may provide a quantitative definition of the degree of ductility; very ductile yielding being characterised by a very small 
$\langle \xi \rangle$, whereas brittle ones would exhibit a large value of $\langle \xi \rangle$. In this view the sharp yielding transition at $\dot{\gamma} \to 0$ corresponds to $\langle \xi \rangle \to \infty$. 

\subsection{Physical interpretation and scaling argument}

We first argue that the typical lengthscale between shear bands, $\langle \xi \rangle$, can be used to assess finite size effects. For $L < \langle \xi \rangle$, a single system-spanning shear band is observed in the simulated system, whereas multiple shear bands appear for $L > \langle \xi \rangle$. 
This reasoning directly explains the system size dependence of the susceptibilities shown in Fig.~\ref{fig:susc}, since stress fluctuations should saturate when the system size becomes larger than $\langle \xi \rangle$. 

Furthermore, the observed scaling behavior, $\langle \xi \rangle \propto \dot{\gamma}^{-\alpha}$, can be physically interpreted as follows. In AQS simulations, the system is given an infinite amount of time to relax in the nearest energy minimum after each strain increment. In practice this energy minimisation takes of course a finite amount of time since the system size is always finite and an efficient conjugate gradient algorithm is employed. Thus, it is possible to observe a unique system-spanning shear band even when the system size increases, $L \to \infty$ (and hence the timescale for shear band formation increases). When a finite shear rate is imposed, however, the shear band only reaches a given size before the external deformation can trigger another shear band elsewhere in the material. Let us define $\tau_{\rm SB}(L)$ the typical timescale for a single shear band to develop inside a system of finite linear size, $L$.  
If we assume that this timescale grows with the system size as 
\begin{equation}
\tau_{\rm SB} \propto L^{\beta},
\label{eq:beta}
\end{equation} 
then it follows that in a very large system deformed at a finite strain rate, the typical distance between the shear bands should scale as 
\begin{equation}
\langle \xi \rangle \propto \dot{\gamma}^{-1/\beta}, 
\label{eq:scaling}
\end{equation}
which suggests a relation between the exponent $\alpha$ introduced in Eq.~(\ref{eq:alpha}) and the exponent $\beta$ characterizing the dynamics of shear band formation in Eq.~(\ref{eq:beta}), namely 
\begin{equation}
\alpha = 1/\beta.
\end{equation} 

To test these ideas, we estimate $\tau_{\rm SB}$ by direct numerical simulations. To this end, we first perform AQS simulations of stable glasses for which a system-spanning shear band forms at yielding accompanying the largest stress drop. In AQS simulations, the stress relaxation is realised by the energy minimisation procedure which uses some unphysical dynamics (such as the conjugate gradient method~\cite{nocedal2006numerical} or FIRE algorithm~\cite{bitzek2006structural}) to reach the energy minimum as quickly as possible. To measure the timescale $\tau_{\rm SB}$, we perform AQS simulations up to the last step before the largest stress drop, but we then switch to the physical steepest descent dynamics described by Eq.~(\ref{eq:shearrate}) at zero strain rate. Thus the system now obeys the physically correct dynamics during the largest stress drop, which allows us to numerically observe the formation of the shear band.

\begin{figure}
\includegraphics[scale=1]{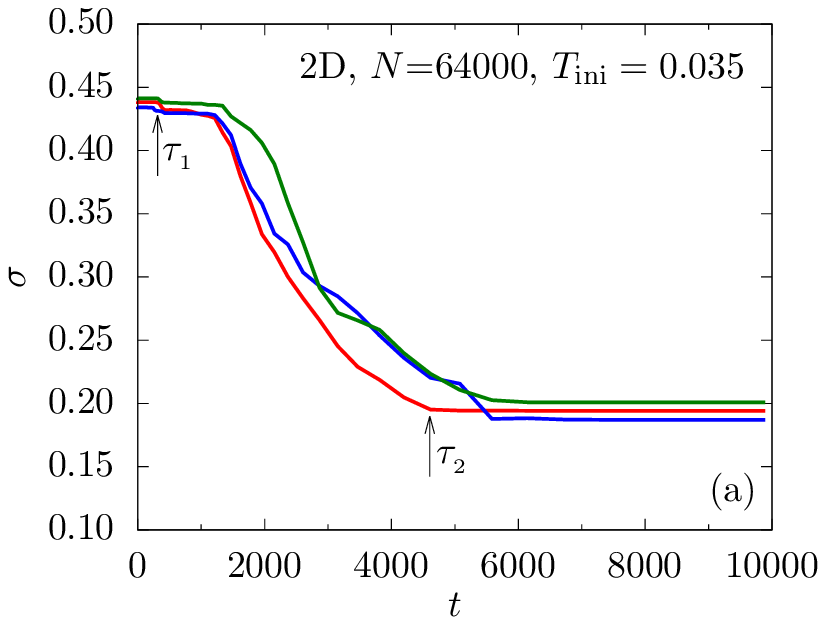}
\includegraphics[scale=1]{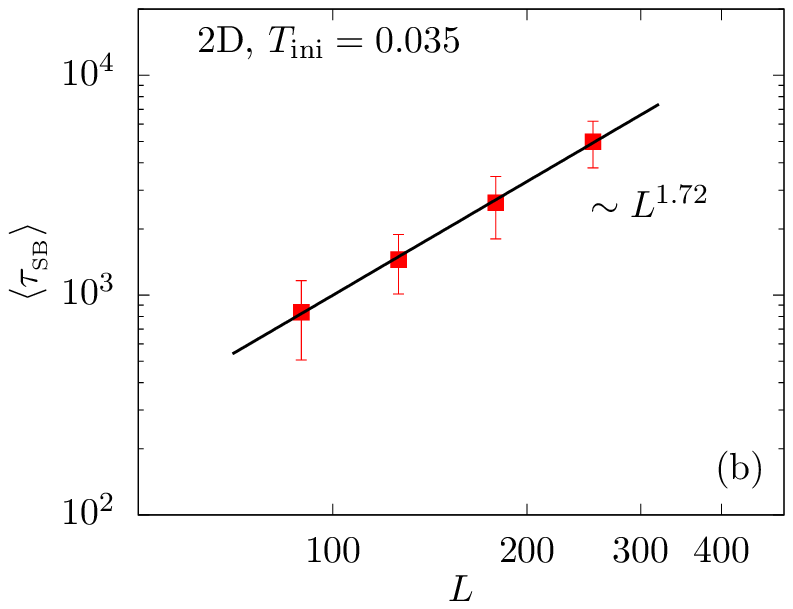}
\caption{(a) Stress relaxation during the largest stress drop for stable glasses using physical steepest descent dynamics. The times $\tau_1$ and $\tau_2$ describe the beginning and the end of the shear band formation for the red curve.
(b) The averaged duration of the shear band formation, $\langle \tau_{\rm SB} \rangle = \langle \tau_2 - \tau_1 \rangle$ increases algebraically with the system size with an exponent $\beta \approx 1.72$ that is then used in Fig.~\ref{fig:lengthXi}.}
\label{fig:relaxation_time}
\end{figure}

In Fig.~\ref{fig:relaxation_time}(a) we show the typical time evolution of the stress, $\sigma(t)$, during the largest stress drop  when the steepest descent dynamics is employed for 2D systems with $N=64000$ and $T_{\rm ini}=0.035$. We show three independent samples. We define $\tau_{\rm SB}=\tau_2-\tau_1$, where $\tau_1$ and $\tau_2$ are the times when $\sigma(t)$ drops $1\%$ below $\sigma(0)$ and when $\sigma(t)$ reaches $1\%$ above $\sigma(\infty)$. These two timescales represent roughly the beginning and the end of the shear band formation, as specified by the arrows in Fig.~\ref{fig:relaxation_time}(a). Therefore, $\tau_{\rm SB}$ quantifies the duration of the formation of a system spanning shear band in a system of finite size $L$. 

We repeat this analysis for many samples (respectively $49$, $68$, $76$, and $73$, for $N=8000$, $16000$, $32000$, and $64000$), from which we deduce the 
average value $\langle \tau_{\rm SB} \rangle$. We then repeat these measurements for various system sizes to estimate how the timescale for the formation of a system spanning shear band grows with the linear size of the system. 
The results are displayed in Fig.~\ref{fig:relaxation_time}(b).
Within the errorbars, we find that $\langle \tau_{\rm SB} \rangle \propto L^{\beta}$ with $\beta \approx 1.72$. This value for the exponent $\beta$ translates into an exponent $\alpha \approx 1/1.72 \approx 0.58$, which is not far from our numerical observations in Fig.~\ref{fig:lengthXi}, although the predicted scaling for $\langle \xi \rangle$ overestimates somewhat the measured growth of the lengthscale. This can be attributed to the fact that the checkerboard patterns in Fig.~\ref{fig:2dsnaps} contain very small shear bands that may bias the chord length distribution towards smaller lengthscales. We expect our prediction to become better when sharp, large shear bands exist, i.e., when both the shear rate $\dot{\gamma}$ and the preparation temperature $T_{\rm ini}$ are small. This trend is compatible with the data shown in Fig.~\ref{fig:lengthXi}.
 
Therefore, our independent analysis supports the idea that the lengthscale $\langle \xi \rangle$ and its scaling with the shear rate result from the competition between the timescale for the formation of a shear band and the imposed shear rate. The degree of brittleness of yielding thus decreases continuously with the imposed shear rate.  

We have repeated the same timescale analysis for the most stable system 
with $T_{\rm ini}=0.062$ in 3D. We again approach the macroscopic stress drop using AQS simulations, but we simulate the largest stress drop dynamics using the physical steepest dynamics. Repeating this analysis for different system sizes, we estimate that the exponent $\beta$ in Eq.~(\ref{eq:beta}) is $\beta \approx 0.96$ in 3D, suggesting that the exponent $\beta$ and hence $\alpha = 1/\beta \approx 1.04$ may depend somewhat on the spatial dimension. Clearly, more work is needed to assess more precisely the value of this exponent, and to understand better the kinetics of the formation of shear bands in amorphous materials since our 2D and 3D data do not allow us to distinguish between diffusive ($\beta = 2$) or ballistic ($\beta=1$) propagation of the shear band. While we may expect that shear bands form ballistically as the macroscopic avalanche unfolds, as we observe here in 3D, our AQS simulations in 2D have revealed the presence of strong spatial disorder-induced fluctuations that may explain the slower shear band propagation (and thus the larger exponent $\beta$) obtained above. We leave this issue for future work.

\section{Discussion and perspectives}

\label{sec:discussions}

We have numerically studied the effect of using a finite strain rate on the  yielding of model glasses prepared over a very wide range of initial stabilities. We found that the non-equilibrium discontinuous transition observed in the AQS conditions for stable glasses~\cite{OBBRT18} is smeared out when a finite deformation rate $\dot{\gamma}$ is imposed. In the quasi-static limit, stable glasses yield at a well-defined yield strain value via the formation of a unique system spanning shear band because the first shear band that appears in the system can propagate throughout the material before the next plastic event occurs. Instead, at finite shear rates, several shear bands can form independently in the material and propagate over a finite lengthscale $\langle \xi \rangle$ that decays algebraically with the shear rate.  
Therefore, the sharp difference between the yielding transitions of poorly-annealed and stable glasses obtained in the quasi-static limit is blurred at finite shear rates where both types of materials display smooth yielding transitions. Yet, the lengthscale $\langle \xi \rangle$ reveals the difference between these two types of materials, since stable systems are characterized by a large distance between localised shear bands (and thus a large value of $\langle \xi \rangle$ that grows when the shear rate is decreased) whereas poorly-annealed glasses display a more homogeneous map of plastic activity (and thus have a small, $\dot{\gamma}$-independent value of $\langle \xi \rangle$). 

Our results open some interesting avenues for future research. It would be interesting to understand and characterise better the lengthscale $\langle \xi \rangle$ in various theoretical settings, from atomistic simulations in various glassy models to more coarse-grained descriptions such as elasto-plastic models where larger system sizes can more easily be studied, in particular perhaps in 3D. More generally, our work should motivate theoretical models, such as soft glassy rheology~\cite{MCF11}, shear transformation zone~\cite{manning2009rate}, elasto-plastic models~\cite{liu2016driving}, mode-coupling theory~\cite{brader2009glass} and random first order transition theory~\cite{wisitsorasak2017dynamical,agoritsas2019out} to address the problem of brittle yielding at finite shear rates.   


Moreover, our results suggest many directions for future computer simulations along the lines proposed here. It would be interesting to study the effect of dimensionality, temperature, and inertia in more details. It would also be important to study other deformation geometries, such as extensional flows to assess the generality of our results. In particular, it has been reported that ductility increases with increasing the shear rate in tensile experiments~\cite{sergueeva2004strain,hajlaoui2008strain} (as we observed numerically), whereas compression tests have demonstrated the opposite trend~\cite{mukai2002effect,antonaglia2014tuned}. More generally, it now becomes possible to simulate the formation of shear band in amorphous solids with stability comparable to the ones of metallic glasses. It thus becomes possible to understand the details of the kinetic mechanism of shear band formation in these systems at the atomic scale.      

\acknowledgments

We thank A. Ninarello for sharing configurations of 3D systems, and J.-L. Barrat, G. Biroli, R. Mari, K. Martens, and G. Tarjus for insightful discussions. This work was supported by a grant from the Simons Foundation (\#454933, L. Berthier).

\bibliography{ref}

\end{document}